\documentclass{article}

\usepackage{arxiv}

\usepackage[utf8]{inputenc} 
\usepackage[T1]{fontenc}    
\usepackage{hyperref}       
\usepackage{url}            
\usepackage{booktabs}       
\usepackage{amsfonts}       
\usepackage{nicefrac}       
\usepackage{microtype}      
\usepackage{lipsum}
\usepackage{graphicx}
\usepackage{amsmath}
\usepackage{graphicx}
\usepackage{float}
\usepackage{inconsolata}
\usepackage[export]{adjustbox} 
\graphicspath{ {./} }

\title{Zero-shot Voice Conversion with Diffusion Transformers}

\author{
 Liu Songting \\
  School of Coumputing and Computer Science\\
  Nanyang Technological University\\
  Singapore \\
  \texttt{lius0114@e.ntu.edu.sg} \\
}

\begin{document}
\maketitle
\begin{abstract}
Zero-shot voice conversion aims to transform a source speech utterance to match the timbre of a reference speech from an unseen speaker. Traditional approaches struggle with timbre leakage, insufficient timbre representation, and mismatches between training and inference tasks. We propose Seed-VC, a novel framework that addresses these issues by introducing an external timbre shifter during training to perturb the source speech timbre, mitigating leakage and aligning training with inference. Additionally, we employ a diffusion transformer that leverages the entire reference speech context, capturing fine-grained timbre features through in-context learning. Experiments demonstrate that Seed-VC outperforms strong baselines like OpenVoice and CosyVoice, achieving higher speaker similarity and lower word error rates in zero-shot voice conversion tasks. We further extend our approach to zero-shot singing voice conversion by incorporating fundamental frequency (F0) conditioning, resulting in comparative performance to current state-of-the-art methods. Our findings highlight the effectiveness of Seed-VC in overcoming core challenges, paving the way for more accurate and versatile voice conversion systems. Code and pretrained models have been release on \texttt{https://github.com/Plachtaa/seed-vc}
\end{abstract}


\section{Introduction}
Voice conversion (VC) technology aims to transform speech from a source speaker to sound as if it were spoken by a target speaker while preserving the original linguistic content. Applications of VC span personalized speech synthesis, dubbing in film and television, and assistance for individuals with speech impairments. Traditional VC methods often require extensive recordings from target speakers, limiting their scalability and practicality.

Zero-shot voice conversion seeks to overcome these limitations by enabling models to convert speech to match the timbre of any unseen speaker, given only a short reference utterance. This capability is crucial for creating flexible and generalizable VC systems that can operate in real-world scenarios with a vast diversity of speakers. Despite significant advancements, zero-shot VC faces several critical challenges.

First, timbre leakage remains a persistent issue. Most existing models \cite{yang2024streamvc}\cite{choi2023diff}\cite{huang2024mullivc}\cite{choi2024dddm} extract speech content from the source utterance using methods like self-supervised learning (SSL) models (e.g., HuBERT, wav2vec)\cite{hsu2021hubert}\cite{baevski2020wav2vec}, automatic speech recognition (ASR) models\cite{zhang2023leveraging}, or phoneme posteriorgrams (PPG)\cite{li2021ppg}. However, these extracted content features often retain residual timbre information from the source speaker, leading to timbre leakage in the converted speech. This leakage causes the converted speech to not fully resemble the timbre of the reference speaker. To mitigate this, some approaches introduce information bottlenecks \cite{qian2019autovc} to filter out timbre information, such as k-means clustering \cite{polyak2021speech} on content features or content retrieval \cite{baas2023voice} based on vector distances. While these methods reduce timbre leakage, they also remove essential content details, resulting in higher word error rates (WER) in the conversion results. This trade-off between speaker similarity and intelligibility is a significant hurdle in zero-shot VC.

Second, insufficient timbre representation poses a challenge. Many prior works represent timbre using a single vector extracted from the reference speech \cite{qian2019autovc}\cite{li2023freevc}. While this may suffice in non-zero-shot scenarios with a limited number of known speakers—where the model parameters can implicitly capture detailed speaker characteristics—it proves inadequate for zero-shot conditions. In zero-shot VC, the model must generalize to unseen speakers with potentially vast differences in vocal attributes. A single-vector representation fails to capture the fine-grained nuances of a speaker's timbre, leading to suboptimal conversion quality.

Third, there is an inconsistency between training and inference tasks. Typically, VC models are trained to reconstruct the source speech using its own content and timbre features. During inference, however, the task shifts to synthesizing speech that combines content from the source utterance with timbre from a different reference speaker. This discrepancy between the training objective and the inference task can degrade performance, as the model is not optimized for the actual conditions encountered during deployment.

To address these challenges, we propose \textbf{Seed-VC}, a novel zero-shot voice conversion framework that introduces two key innovations. First, we employ an external timbre shifter during training to perturb the timbre of the source speech. The timbre shifter can be a non-perfect zero-shot VC model or a semantic-to-acoustic model from an existing text-to-speech (TTS) system. By transforming the source speech into a timbre-shifted version, we obtain an alternative utterance from which we extract the content features. This process ensures that the content extractor operates on speech that does not carry the original timbre, effectively reducing timbre leakage. Moreover, this strategy aligns the training procedure with the inference scenario, where content and timbre come from different speakers.

Second, we utilize a diffusion transformer \cite{peebles2023scalable} architecture capable of leveraging the entire reference speech as context rather than relying on a single timbre vector. The transformer's in-context learning ability allows the model to capture more detailed and nuanced timbre information from the reference speech. In-context learning ability of transformers has been proved to largely increase speaker similarity in many zero-shot TTS models\cite{wang2023neural}\cite{lee2024ditto}\cite{eskimez2024e2}\cite{chen2024f5}, but seldom used in VC task. By incorporating the full reference utterance, the model can learn fine-grained speaker characteristics essential for high-quality zero-shot conversion.

Our experimental results demonstrate that Seed-VC significantly outperforms strong baselines like OpenVoice \cite{qin2023openvoice} and CosyVoice\cite{du2024cosyvoice}, achieving higher speaker similarity and lower WER in zero-shot voice conversion tasks. We also compare our model's zero-shot performance with previous non-zero-shot models on several speakers, achieving equivalent or superior speaker similarity with much lower WER.

Additionally, we extend our framework to zero-shot singing voice conversion by incorporating fundamental frequency (F0) conditioning into the model inputs. The results show high speaker similarity and low WER while maintaining high F0 correlation and low F0 root mean square error (RMSE), as well as high DNSMOS scores, indicating naturalness and intelligibility.

In summary, our contributions are as follows:

\begin{itemize}
    \item We propose \textbf{Seed-VC}, a novel framework that addresses timbre leakage and training-inference inconsistency by using an external timbre shifter during training.
    \item We enhance timbre representation through a diffusion transformer that utilizes the entire reference speech context, capturing fine-grained speaker characteristics crucial for zero-shot VC.
    \item We demonstrate superior performance over existing state-of-the-art models in both zero-shot voice conversion and zero-shot singing voice conversion tasks.
    \item We provide insights through ablation studies on the impact of the timbre shifter, the use of full reference speech context, and different timbre shifter methods.
\end{itemize}
Our work advances the field of zero-shot voice conversion by addressing core challenges and providing a more effective and generalizable approach. We believe that Seed-VC paves the way for more accurate and versatile voice conversion systems applicable to a wide range of real-world scenarios.

\section{Related works}
In this section, we review prior work relevant to our proposed Seed-VC framework, focusing on content representation, timbre modeling, and techniques to mitigate timbre leakage and training-inference discrepancies.
\subsection{Content Representation in Voice Conversion}
Accurate extraction of linguistic content from speech is crucial for VC systems. Traditional methods often rely on parallel data and explicit alignment techniques, which are not scalable for zero-shot scenarios. Self-supervised learning (SSL) models have emerged as powerful tools for content representation without the need for labeled data.

\paragraph{Self-Supervised Models} SSL models like HuBERT\cite{hsu2021hubert} and wav2vec 2.0\cite{baevski2020wav2vec} learn robust speech representations by predicting masked portions of the input signal. These models capture phonetic and linguistic information while being less sensitive to speaker-specific characteristics, making them suitable for extracting content features in VC.

\paragraph{Phonetic Posteriorgrams (PPGs)} PPGs\cite{sun2016phonetic} are derived from automatic speech recognition (ASR) models and represent the posterior probabilities of phonetic units. They have been widely used in VC for content representation. However, PPGs can retain residual speaker information, leading to timbre leakage in the converted speech.

\subsection{Addressing Timbre Leakage}
Timbre leakage occurs when residual speaker information from the source speech content representation contaminates the converted speech, affecting the timbre similarity with the target speaker. Several approaches have been proposed to mitigate this issue.

\paragraph{Information Bottleneck Methods} Introducing a bottleneck in the content representation can reduce the amount of speaker information retained. For instance, AutoVC\cite{qian2019autovc} uses a carefully designed bottleneck in the encoder to separate content from speaker characteristics. However, aggressive bottlenecking can lead to loss of important linguistic information, increasing the word error rate (WER) in the converted speech.

\paragraph{Discretization Techniques} Methods like vector quantization (VQ) and k-means clustering have been employed to discretize content features, effectively removing speaker variability. VQ-VAE models\cite{van2017neural} quantize the latent space to encourage disentanglement of content and timbre. While these methods reduce timbre leakage, they may also discard subtle linguistic details essential for natural-sounding speech.

\paragraph{Content Retrieval Based on Vector Distances} Some approaches retrieve content representations that are less correlated with timbre by selecting vectors based on distance metrics\cite{baas2023voice}. This can help in reducing timbre leakage but may introduce distortions in content representation, affecting speech intelligibility.

\subsection{Diffusion Models and Diffusion Transformers in Speech Processing}
Recent advancements in generative modeling have introduced diffusion probabilistic models as a powerful approach for data generation tasks. Originally proposed for image synthesis\cite{ho2020denoising}, diffusion models have demonstrated remarkable capabilities in modeling complex data distributions through a sequential denoising process. This methodology has been extended to the speech domain, yielding significant improvements in speech synthesis and voice conversion.

\paragraph{Diffusion Models in Speech Synthesis} Models like DiffWave\cite{kong2020diffwave} and WaveGrad\cite{chen2020wavegrad} apply the diffusion process to waveform generation, achieving high-fidelity audio synthesis with reduced computational requirements compared to traditional autoregressive models. These models iteratively refine a noise signal towards a target speech waveform, guided by learned denoising functions. The success of diffusion models in speech synthesis highlights their potential for capturing intricate temporal and spectral patterns in audio data.

\paragraph{Diffusion Models in Voice Conversion} In voice conversion, diffusion models have been explored to enhance the naturalness and quality of the converted speech. By modeling the data distribution of speech signals, diffusion-based VC models can generate outputs that closely adhere to the characteristics of the target speaker while preserving the linguistic content. For instance, diffusion probabilistic modeling has been used to improve the robustness of VC systems against over-smoothing and to better handle the variability in speech data\cite{popov2021diffusion}\cite{choi2023diff}\cite{choi2024dddm}.

\paragraph{Diffusion Transformers} Combining diffusion models with transformer architectures leverages the strengths of both approaches. Transformers\cite{vaswani2017attention}, known for their powerful sequence modeling and in-context learning capabilities, are adept at capturing long-range dependencies in sequential data. Integrating transformers into the diffusion framework allows for more expressive modeling of speech signals.
In the context of text-to-speech (TTS), models like DiffSinger\cite{liu2022diffsinger} employ diffusion transformers to generate high-quality singing voices by modeling the distribution of mel-spectrograms conditioned on phonetic and pitch information. The in-context learning ability of transformers enables the model to utilize the entire sequence of reference speech, capturing fine-grained nuances essential for natural-sounding synthesis.

\paragraph{Application to Zero-Shot Voice Conversion} For zero-shot VC, diffusion transformers offer the advantage of handling variable-length sequences and incorporating rich contextual information from the reference speech. By conditioning on the full reference utterance, the model can learn detailed timbre representations necessary for accurately mimicking unseen speakers. This approach addresses the limitations of single-vector timbre representations and enhances the model's ability to generalize in zero-shot scenarios.

Our proposed Seed-VC framework builds upon these advancements by adopting a diffusion transformer architecture. By leveraging the diffusion process within a transformer model, Seed-VC aims to capture more nuanced timbre characteristics from the reference speech through in-context learning. This integration enables the model to produce high-quality, natural-sounding converted speech that closely matches the timbre of unseen speakers, addressing the challenges identified in prior works.

\section{Proposed Method}
In this section, we introduce Seed-VC, our proposed framework for zero-shot voice conversion that addresses the challenges of timbre leakage, insufficient timbre representation, and training-inference inconsistency. Seed-VC leverages a diffusion transformer architecture and employs an external timbre shifter during training to enhance performance in zero-shot scenarios.

\begin{figure}[H]
\centering
\includegraphics[center,width=0.6\textwidth]{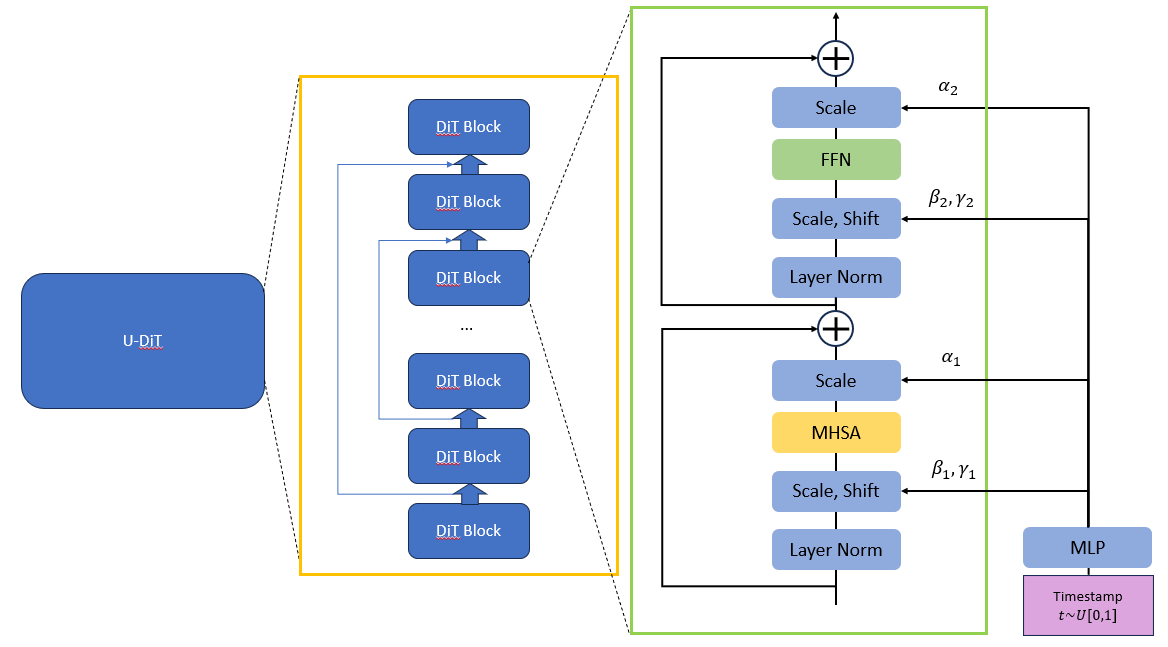}
\caption{\textbf{Architectural Detail} about U-Net style skip connections and timestamp condition.}
\label{architecture}
\end{figure}

\subsection{Flow Matching}
We adopt a flow matching diffusion scheme for training and inference. The flow matching algorithm aligns distributions by minimizing the discrepancy between a source flow $\mathbf{f}_s$ and a target flow $\mathbf{f}_t$. Given a source distribution $p_s(\mathbf{x})$ and a target distribution $p_t(\mathbf{x})$, the algorithm constructs a continuous path by learning a time-dependent vector field $\mathbf{v}(\mathbf{x}, t)$ that gradually morphs $p_s$ into $p_t$.

The transformation is guided by minimizing the L1 loss:
\begin{equation}
    \mathcal{L}_{\text{FM}} = \mathbb{E}_{\mathbf{x} \sim p_s, t \sim U[0,1]} \left[ \left\| \mathbf{f}_s(\mathbf{x}, t) - \mathbf{v}(\mathbf{x}, t) \right\| \right]\label{fm loss}
\end{equation}
\begin{equation}
\mathbf{v}(\mathbf{x}, t)=\mathbf{f}_\text{estimator}(\mathbf{x}, t, \textbf{c})\label{fm condition}
\end{equation}
where $\mathbf{f}_s(\mathbf{x}, t)$ represents the actual velocity of $\mathbf{x}$ at time $t$ and $\mathbf{v}(\mathbf{x}, t)$ denotes the estimated vector field predicted by the model $\mathbf{f}_\text{estimator}$ conditioned on context $\mathbf{c}$. The optimization objective ensures that the learned flow $\mathbf{v}(\mathbf{x}, t)$ aligns closely with the target distribution over time. This alignment is typically achieved by solving an ordinary differential equation (ODE):
\begin{equation}
\frac{d\mathbf{x}(t)}{dt} = \mathbf{v}(\mathbf{x}(t), t), \quad \mathbf{x}(0) \sim p_s.
\end{equation}
The solution to this ODE provides a path from the source to the target distribution. 

\subsection{Model Architecture}
The main components of the framework are:

\begin{itemize}
\item Diffusion Transformer: a transformer-based network with $N$ layers of dimension $d$, which models the denoising process in the diffusion scheme. We further apply improvements from U-ViT\cite{bao2023all}:
\begin{itemize}
    \item \textbf{U-Net style skip connection: } We apply skip connections following similar method as U-Net \cite{ronneberger2015u}. However, we do not apply down-sampling to sequence length but keep it consistent through model forward. A detailed illustration is presented in Figure \ref{architecture}
    \item \textbf{Time as token: } We prepend time embedding as a prefix token in the sequence. Meanwhile, the same time embedding is also used as adaptive layernorm in transformer blocks, similar to \cite{peebles2023scalable}.
    \item \textbf{Rotary positional embedding: } We apply rotary positional embedding following \cite{su2023enhanced}, which gives better generalization performance of position encoding and has certain extrapolation performance
\end{itemize}
\item Length Regulator: a convolution stack. since semantic features may not have the same frame rate as the acoustic feature, we introduce a length regulator module to interpolate the original semantic feature sequence to the same length as the acoustic feature. Specifically, semantic feature sequence is first nearest-interpolated to a desired length (in training, it is the acoustic feature sequence length), then passed to a convolution stack to get a smoother representation of speech content signal.
\end{itemize}
The input to the diffusion transformer at each timestep $t$ consists of:

\begin{itemize}
    \item Timbre Vector: $e_{\text{timbre}} \in \mathbb{R}^d$, where $d$ is the embedding dimension. This is extracted by a pretrained external speaker verification model.
    \item Semantic Features: $\mathbf{S} = [\mathbf{s}_1, \mathbf{s}_2, \dots, \mathbf{s}_L]$, where $\mathbf{s}_i \in \mathbb{R}^d$, and $L$ is the sequence length after interpolation.
    \item Noisy Acoustic Features: $\tilde{\mathbf{A}}_t = [\tilde{\mathbf{a}}_1^t, \tilde{\mathbf{a}}_2^t, \dots, \tilde{\mathbf{a}}_L^t]$, where $\tilde{\mathbf{a}}_i^t$ is the noisy acoustic feature at timestep $t$, and $L$ is the length of the acoustic feature sequence.
    \item Time step: $t \in [0, 1]$
\end{itemize}

During training, a random portion of the target acoustic feature sequence is selected as the audio prompt, while the remaining portion is replaced with noise and serves as the target for denoising. The semantic features for the entire utterance are provided without masking. The timbre vector is perpended as a prefix to the transformer input to inform the model about the target speaker's timbre. The diffusion timestamp information is incorporated as both prefix token and adaptive layer normalization within the transformer blocks. Hence, context $\mathbf{c}$ in \eqref{fm condition} is formulated as:

\begin{equation}
    \mathbf{c}=[e_{timbre}, \mathbf{S}]
\end{equation}
The model predicts the vector field $\mathbf{v}(\mathbf{x}, t)$ conditioned on $\mathbf{c}$ following \eqref{fm loss}, where:
\begin{equation}
    \mathbf{x}=\tilde{\mathbf{A}}_t
\end{equation}

\begin{figure}[H]
\centering
\includegraphics[center,width=0.6\textwidth]{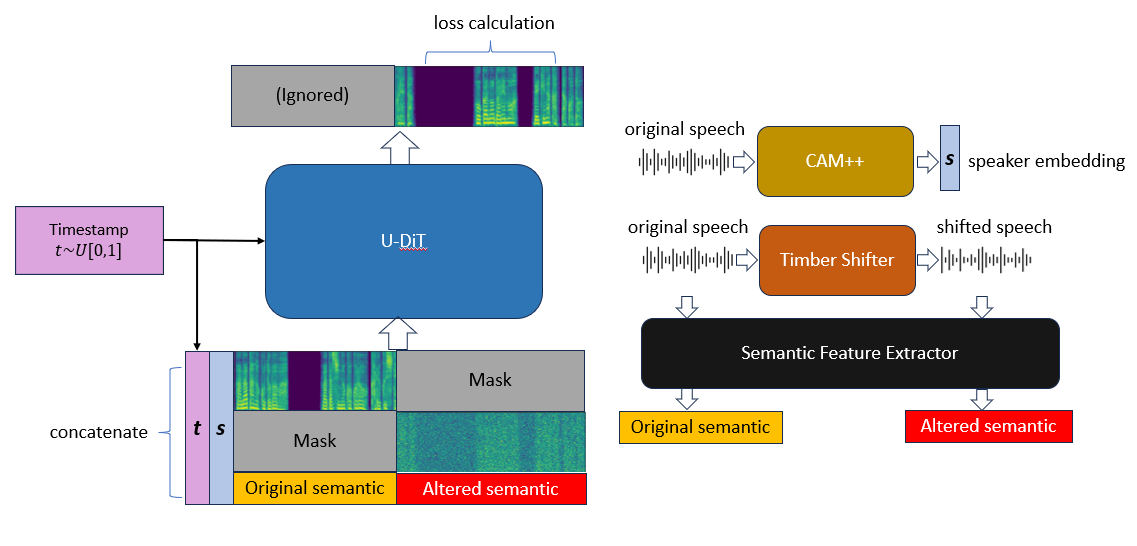}
\caption{\textbf{Training Pipeline.} A random segment is set as timbre prompt. Prompt component contains semantic feature and acoustic feature from original audio, while target component contains semantic feature from timbre-shifted audio. Loss is only calculated on target component.}
\label{train}
\end{figure}

\subsection{Addressing Timbre Leakage and Training-Inference Gap}
To mitigate timbre leakage and align the training process with the inference scenario, we introduce an external timbre shifter module during training. The timbre shifter alters the timbre of the source utterance, ensuring that the semantic features used as input do not contain residual timbre information from the source speaker.

\subsubsection{Timbre Shifter Module}
The timbre shifter $\mathcal{T}$ transforms the source utterance $\mathbf{X}_{\text{src}}$ into an utterance with a randomly altered timbre $e_{r}$:

\begin{equation}\label{timbre shift}
    \mathbf{X}_{\text{shifted}}=\mathcal{T}(\mathbf{X}_{\text{src}}, e_{r})
\end{equation}
where $e_{r}$ is a timbre representation recognizable by the external timbre shifter.  

We explore two types of timbre shifters:

\begin{itemize}
    \item Zero-Shot Voice Conversion Models: Models like AutoVC\cite{qian2019autovc}, FreeVC\cite{li2023freevc}, or YourTTS\cite{casanova2022yourtts}, which can convert speech to a target timbre, albeit imperfectly. Using these models as timbre shifters changes the timbre of the source utterance sufficiently for our purposes.
    \item Semantic-to-Acoustic Modules from TTS Models: The second stage of TTS models (e.g., CosyVoice\cite{du2024cosyvoice}'s diffusion-based module) converts semantic tokens to acoustic features and inherently changes the timbre when provided with different reference audios.
\end{itemize}

\subsubsection{Training Process with Timbre Shifter}\label{train with timbre shifter}
As depicted in Figure \ref{train}, training process involves the following steps:

\begin{enumerate}
    \item \textbf{Timbre Shifting}:  
    Apply the timbre shifter to the source utterance as \eqref{timbre shift}.
    \item \textbf{Semantic Feature Extraction}:
    Extract semantic features from both the source and shifted utterances using semantic encoder $\mathbf{f}_\text{semantic}$:
    \begin{equation}
        \mathbf{S}_\text{src}=\mathbf{f}_\text{semantic}(\mathbf{X}_{\text{src}})
    \end{equation}
    \begin{equation}
        \mathbf{S}_\text{shifted}=\mathbf{f}_\text{semantic}(\mathbf{X}_{\text{shifted}})
    \end{equation}
    \item \textbf{Acoustic Feature Preparation}:
    Extract acoustic features (mel spectrogram) from the source utterance waveform with pre-defined transformation $\mathbf{f}_\text{mel}$:
    \begin{equation}
        \mathbf{A}=\mathbf{f}_\text{mel}(\mathbf{X}_{\text{src}})
    \end{equation}
    \item \textbf{Prompt and Noise Components}:
    \begin{itemize}
        \item Randomly select a segment of the acoustic feature $\mathbf{A}$ as the audio prompt $\mathbf{A}_{\text{prompt}}$.
        \item The remaining portion is used as the target for denoising, with added noise corresponding to timestep $t$.
    \end{itemize}
    \item \textbf{Model Input Construction}: The model input consists of:
    \begin{itemize}
        \item Timbre vector $e_{\text{timbre}}$ extracted from the source utterance using external speaker verification model $\textbf{f}_\text{sv}$:
        \begin{equation}
            \mathbf{e}_\text{timbre}=\textbf{f}_\text{sv}(\mathbf{X}_{\text{src}})
        \end{equation}
        \item Semantic features:
        \begin{itemize}
            \item For the prompt component: $\mathbf{S}_{\text{src}}$
            \item For the noise component: $\mathbf{S}_{\text{shifted}}$
        \end{itemize}
         \item Acoustic features:
        \begin{itemize}
            \item For the prompt component: original acoustic features $\mathbf{A}$.
            \item For the noise component: noisy acoustic features $\tilde{\mathbf{A}}_t$.
        \end{itemize}
    \end{itemize}
\end{enumerate}

\begin{figure}[H]
\centering
\includegraphics[center,width=0.6\textwidth]{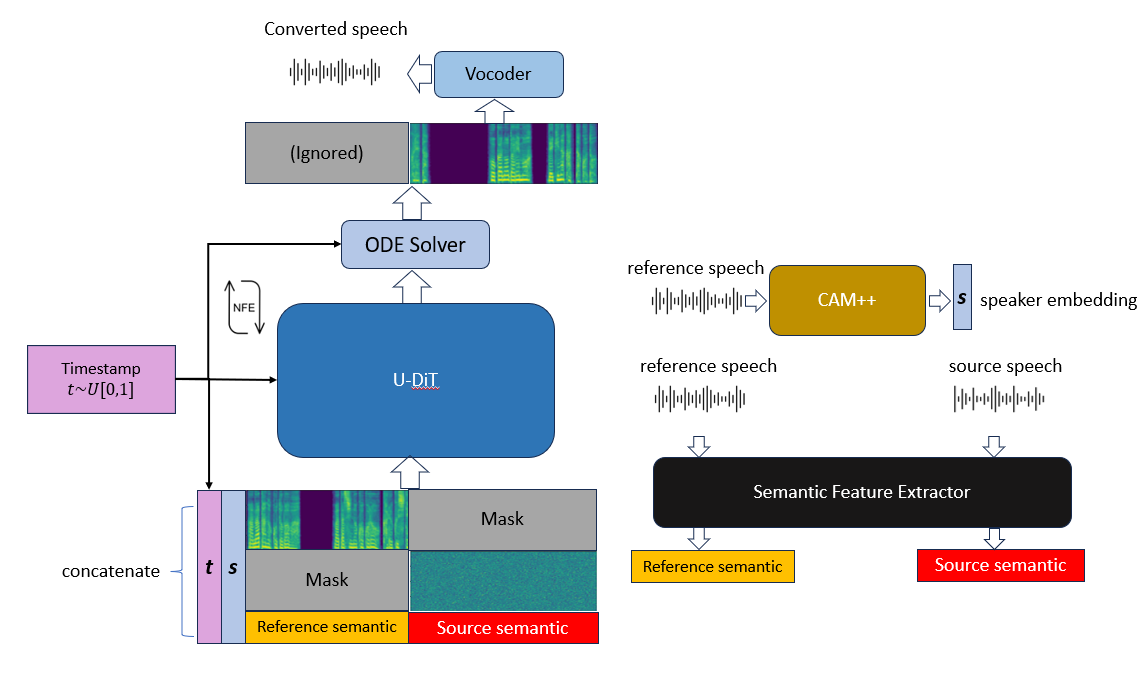}
\caption{\textbf{Inference Pipeline.} Corresponds to training pipeline, reference audio plays the role as timbre enrollment.} 
\label{inference}
\end{figure}

\subsection{Enhanced Timbre Representation}
As in Section \ref{train with timbre shifter}, a random audio segment is used as context and excluded from loss calculation. This allows the model to capture more fine-grained timbre information besides timbre vector. During inference process shown in Figure \ref{inference}, the entire reference audio can be included as model input. We will show how this feature boost performance in aspect of timbre similarity in Section \ref{experiments}

\subsection{Extension to Zero-Shot Singing Voice Conversion}
Singing voice conversion introduces additional challenges due to the expressive nature of singing and the importance of pitch and prosody. To extend Seed-VC to zero-shot singing voice conversion, we incorporate fundamental frequency (F0) conditioning into the model.

We process the F0 information as follows:
\begin{enumerate}
    \item \textbf{F0 Extraction}: Extract the F0 contour from the source singing voice using a pitch estimation algorithm.
    \item \textbf{Quantization}: Exponentially quantize the continuous F0 values into 256 bins:
    \begin{equation}
        q_{F0}(f)=\text{Quantize}(log(f))
    \end{equation}
    This quantized F0 representation captures the pitch contour in a form suitable for conditioning.
    \item Conditioning the Model: Add the quantized F0 information to the semantic features as additional conditioning:
    \begin{equation}
        \mathbf{S}'=\text{Concat}(\mathbf{S}, q_{F0}(\mathbf{F}))
    \end{equation}
    Where $\mathbf{F}$ is the F0 contour, and $\mathbf{S}'$ is the augmented semantic feature sequence.
    \item Gender-Based Pitch Shift Adjustment
    \begin{itemize}
        \item During inference, we adjust the F0 contour based on the gender of the source and target speakers to account for typical pitch ranges:
        \begin{itemize}
            \item If the source and target genders differ, apply a pitch shift to align the F0 contour appropriately. Specifically, we apply +12 semitones for male-to-female conversion and -12 semitones for female-to-male conversion.
        \end{itemize}
    \end{itemize}
\end{enumerate}

By incorporating F0 conditioning, the model can capture the pitch variations essential for singing. The adjusted semantic features $\mathbf{S}'$ provide the model with both linguistic and pitch information, enabling accurate conversion of singing voices while preserving the melodic and expressive qualities of the original performance.

\section{Experiments}\label{experiments}
In this section, we evaluate the performance of the proposed Seed-VC framework on zero-shot voice conversion and zero-shot singing voice conversion tasks. We compare our model with strong baseline systems and analyze the results using various objective metrics.
\subsection{Experimental Setup}
\subsubsection{Model Configuration}
\paragraph{Training} The Seed-VC base model has 13 layers, 8 attention heads, 512/2048 embedding/feed-forward network (FFN) dimension for DiT. The model predicts mel spectrogram of 22.05kHz sampling rate, 1024 Fast Fourier Transform (FFT) window and 256 hop size, with 80 mel bins. For singing voice conversion, we trained a larger version with 17 layers, 12 attention heads, 768/3072 embeeding/FFN dimension. The sampling rate is 44.1kHz for better audio quality, 2048 FFT window and 512 hop size, with 128 mel bins. Both models are trained with equivalent batch size of 64 for 400k steps. The optimizer is AdamW with peak learning rate of 1e-4, exponentially decays to a minimum of 1e-5.
\paragraph{Timbre Shifter} For the timbre shifter module used during training (as described in Section 3.2), we employed OpenVoiceV2\cite{qin2023openvoice}, which is essentially a YourTTS\cite{casanova2022yourtts} model trained on a proprietary large-scale dataset. Although the exact details of the dataset are not publicly available, OpenVoice effectively alters the timbre of source utterances, making it suitable for our purposes. While we considered training a YourTTS model from scratch as a substitute, we opted to use OpenVoiceV2 to conserve computational resources and time.
\paragraph{Semantic Encoder} We use the encoder of Whisper-small\footnote{https://github.com/openai/whisper}\cite{radford2023robust} from OpenAI as the semantic encoder. As an end-to-end encoder-decoder ASR model, Whisper's encoder is good at capturing linguistic content and discarding speaker-related information. We do not perform any discretization techniques such as K-means or vector quantize (VQ), as it is very likely to lose detailed linguistic information if not properly trained.
\paragraph{Speaker Encoder} We timbre vector using a pretrained speaker verification model CAM++\footnote{https://github.com/modelscope/3D-Speaker/tree/main/egs/3dspeaker/sv-cam} \cite{wang2023cam++}. It was trained on a large scale dataset with over 200k speakers, enabling its strong timbre generalization ablility.
\paragraph{Vocoder} For transformation from mel-spectrogram to waveform, we use pretrained BigVGANV2\footnote{https://github.com/NVIDIA/BigVGAN}\cite{lee2022bigvgan} model from NVIDIA.
\paragraph{F0 extractor} For singing voice conversion, the additional F0 input is extracted by a pretrained RMVPE\footnote{https://github.com/Dream-High/RMVPE}\cite{wei2023rmvpe} model.
\subsubsection{Datasets}
\paragraph{Training Dataset} We trained Seed-VC using the Emilia-101k\footnote{https://huggingface.co/datasets/amphion/Emilia-Dataset}\cite{he2024emilia} dataset, which consists of approximately 101,000 hours of speech data covering a wide range of speaking styles and content. This dataset was compiled from publicly available sources on the internet, providing diverse linguistic and acoustic variations that are crucial for training a robust zero-shot voice conversion model.
\paragraph{Evaluation Dataset}
For evaluating zero-shot voice conversion performance, we randomly selected 100 utterances from the LibriTTS test-clean dataset as source speech samples. The LibriTTS dataset is widely used for speech synthesis and voice conversion research due to its high-quality recordings and diverse speaker pool. As target speakers, we selected 8 random utterances from the seed-tts-eval dataset, ensuring that both source and target speakers were unseen during training to align with the zero-shot conversion setting.
\subsubsection{Evaluation Metrics}
To comprehensively assess the performance of Seed-VC and baseline models, we used the following objective metrics:
\begin{itemize}
    \item \textbf{Speaker Similarity (SECS):} Measured as the cosine similarity between speaker embeddings extracted from the converted speech and the reference target speech using a speaker verification model. A higher SECS indicates better timbre similarity to the target speaker. In this work, we use Resemblyzer\footnote{https://github.com/resemble-ai/Resemblyzer} to extract speaker embeddings.
    \item \textbf{Word Error Rate (WER):} Calculated by comparing the transcriptions of the converted speech (obtained via an automatic speech recognition system, in this work we use HuBERT\footnote{https://huggingface.co/facebook/hubert-large-ls960-ft}\cite{hsu2021hubert} finetuned for ASR) with the ground truth transcripts. Lower WER values signify better preservation of linguistic content.
    \item \textbf{Character Error Rate (CER):} Similar to WER but computed at the character level, providing a finer-grained evaluation of transcription accuracy.
    \item \textbf{DNSMOS P.835 Scores:}Predicted using Microsoft's DNSMOS P.835 model\cite{reddy2021dnsmos}, which provides non-intrusive objective speech quality scores across three dimensions:
    \begin{itemize}
        \item \textbf{SIG (Signal Distortion):} Reflects speech signal quality.
        \item \textbf{BAK (Background Noise Intrusiveness):} Assesses the level of background noise.
        \item \textbf{OVRL (Overall Quality):} An overall quality score combining SIG and BAK. Higher scores indicate better perceived speech quality.
    \end{itemize}
\end{itemize}
\subsection{Baseline Models}
We compared Seed-VC against two state-of-the-art, open-source baseline models:
\begin{itemize}
    \item \textbf{OpenVoice:\footnote{https://github.com/myshell-ai/OpenVoice}\cite{qin2023openvoice}} An implementation based on YourTTS, trained on a proprietary large-scale dataset. OpenVoice is designed for zero-shot voice conversion and has shown robust performance across various speakers and linguistic contexts.
    \item \textbf{CosyVoice:\footnote{https://github.com/FunAudioLLM/CosyVoice}\cite{du2024cosyvoice}} An open-source text-to-speech (TTS) model released by Alibaba. While being a TTS model, it has also published a variant for voice conversion. CosyVoice employs a diffusion-based semantic-to-acoustic module, enabling it to perform zero-shot voice conversion by leveraging its powerful semantic and acoustic modeling capabilities.
\end{itemize}
Both baseline models were selected due to their strong performance in zero-shot voice conversion tasks and their use of large-scale training datasets, providing a challenging benchmark for evaluating Seed-VC.
\subsection{Results and Discussion}
\subsubsection{Zero-Shot Voice Conversion Performance}
We evaluated the zero-shot voice conversion performance of Seed-VC and the baseline models using the metrics described above. The results are presented in \textbf{Table 1}.
\begin{table}[h]
\centering
\begin{tabular}{|l|c|c|c|c|c|c|}
\hline
\textbf{Models} & \textbf{SECS↑} & \textbf{WER↓} & \textbf{CER↓} & \textbf{SIG↑} & \textbf{BAK↑} & \textbf{OVRL↑} \\
\hline
Ground Truth & 1.0000 & 8.02 & 1.57 & -- & -- & -- \\
OpenVoice & 0.7547 & 15.46 & 4.73 & \textbf{3.56} & \textbf{4.02} & \textbf{3.27} \\
CosyVoice & 0.8440 & 18.98 & 7.29 & 3.51 & 4.02 & 3.21 \\
Seed-VC (without full reference enrollment) & 0.7948 & \textbf{11.98} & \textbf{3.03} & 3.37 & 3.93 & 3.05 \\
Seed-VC (Ours) & \textbf{0.8676} & \textbf{11.99} & \textbf{2.92} & 3.42 & 3.97 & 3.11 \\
\hline
\end{tabular}
\caption{Zero-Shot Voice Conversion Performance. \textit{Note: Higher values are better for metrics with an upward arrow (↑), and lower values are better for metrics with a downward arrow (↓). The Ground Truth represents the original unmodified speech.}}
\end{table}

\paragraph{Speaker Similarity (SECS)} Seed-VC achieved the highest speaker similarity score of 0.8676, outperforming both OpenVoice and CosyVoice. This indicates that Seed-VC is more effective at capturing and reproducing the target speaker's timbre in zero-shot scenarios, validating the effectiveness of our enhanced timbre representation and training strategies. We also observed severe degradation in speaker similarity if only timbre vector is given while full reference audio is not enrolled as context, highlighting the importance of in-context learning in voice conversion task.

\paragraph{Word Error Rate (WER) and Character Error Rate (CER)} Seed-VC also achieved the lowest WER (11.99\%) and CER (2.92\%), demonstrating superior preservation of linguistic content during the conversion process. Unlike CosyVoice, we input continous semantic feature instead of discrete tokens, avoiding potential loss in speech content. Meanwhile, with our novel training strategy with timbre shifter, the timbre leakage issue appeared in continuous semantic feature is avoided. This suggests that our approach effectively mitigates the trade-off between timbre similarity and speech intelligibility commonly observed in zero-shot voice conversion.

\paragraph{DNSMOS P.835 Scores} In terms of perceived speech quality, OpenVoice achieved slightly higher scores in SIG, BAK, and OVRL dimensions. However, the differences are marginal, and Seed-VC's performance remains competitive. The slightly lower SIG and OVRL scores for Seed-VC may be attributed to the trade-off between enhancing timbre similarity and maintaining speech quality.

The superior performance of Seed-VC in speaker similarity and intelligibility metrics highlights the effectiveness of incorporating the external timbre shifter and the diffusion transformer architecture. By addressing timbre leakage and aligning the training and inference processes, Seed-VC better generalizes to unseen speakers without sacrificing content accuracy.

\subsubsection{Zero-Shot Singing Voice Conversion}
To evaluate the applicability of Seed-VC to zero-shot singing voice conversion, we conducted experiments using the M4Singer\footnote{https://github.com/M4Singer/M4Singer}\cite{zhang2022m4singer} dataset, which contains high-quality singing recordings. We selected four target singers from an in-the-wild speech dataset\footnote{https://huggingface.co/datasets/XzJosh/audiodataset}, ensuring diversity in vocal characteristics and styles.  

\textbf{Experimental Setup}
\begin{itemize}
    \item \textbf{Source Audio:} We used 100 random singing utterances for each singer type as source audio.
    \item \textbf{Target Speakers (Characters):} For each character, one random utterance was chosen as the prompt for zero-shot inference.
    \item \textbf{Baseline Model:} We trained a respective RVCv2-f0-48k model for each character as a baseline. RVCv2\footnote{https://github.com/RVC-Project/Retrieval-based-Voice-Conversion-WebUI} is a popular open-source project 
 based on VITS\cite{kim2021conditional}, it has demonstrated strong performance in singing voice conversion tasks.
    \item \textbf{Additional Evaluation Metrics}
    \begin{itemize}
        \item \textbf{F0 Correlation (F0CORR):} Measures the correlation between the F0 contours of the converted and source singing voices. Higher values indicate better pitch contour preservation.
        \item \textbf{F0 Root Mean Square Error (F0RMSE):}Evaluates the deviation in pitch between the converted and source singing voices. Lower values are better.
    \end{itemize}
\end{itemize}

\begin{table}[h]
\centering
\begin{tabular}{|l|c|c|c|c|c|c|c|}
\hline
\textbf{Models} & \textbf{FOCORR↑} & \textbf{FORMSE↓} & \textbf{SECS↑} & \textbf{CER↓} & \textbf{SIG↑} & \textbf{BAK↑} & \textbf{OVRL↑} \\
\hline
RVCv2 & \textbf{0.9404} & 30.43 & 0.7264 & 28.46 & \textbf{3.41} & \textbf{4.05} & \textbf{3.12} \\
Seed-VC (Ours) & 0.9375 & \textbf{33.35} & \textbf{0.7405} & \textbf{19.70} & 3.39 & 3.96 & 3.06 \\
\hline
\end{tabular}
\caption{Zero-Shot Singing Voice Conversion Performance}
\end{table}

\paragraph{Speaker Similarity (SECS):} Seed-VC achieved a higher speaker similarity score (0.7405) compared to RVC v2 (0.7264), indicating better timbre adaptation in the context of singing voices.

\paragraph{F0CORR and F0RMSE:} Both models achieved high F0 correlation coefficients (approximately 0.94), demonstrating accurate preservation of pitch contours essential for singing voice conversion. RVC v2 had a slightly lower F0RMSE, suggesting marginally better pitch accuracy.

\paragraph{Character Error Rate (CER):} Seed-VC significantly outperformed RVC v2 in CER, achieving a lower error rate (19.70\% vs. 28.46\%). This indicates that Seed-VC more effectively preserves the lyrical content during singing voice conversion, which is crucial for intelligibility in songs. Note that since M4Singer is in Chinese, we use SenseVoice\footnote{https://github.com/FunAudioLLM/SenseVoice}\cite{speechteam2024funaudiollm} as the ASR model in this evaluation.

\paragraph{DNSMOS P.835 Scores:} RVCv2 obtained slightly higher DNSMOS scores in SIG, BAK, and OVRL dimensions. However, the differences are minimal, and Seed-VC's performance remains competitive in terms of perceived audio quality.

The results demonstrate that Seed-VC effectively extends to zero-shot singing voice conversion, maintaining high speaker similarity and better preservation of lyrical content compared to the baseline. The incorporation of F0 conditioning and gender-based pitch shift adjustments (as detailed in Section 3.4) enables the model to handle the expressive and melodic nature of singing voices. While there is a slight trade-off in pitch accuracy and speech quality, the overall performance indicates that Seed-VC is a viable solution for zero-shot singing voice conversion tasks.

\section{Conclusion}
In this paper, we introduced \textbf{Seed-VC}, a novel zero-shot voice conversion framework designed to address the key challenges of timbre leakage, insufficient timbre representation, and training-inference inconsistency prevalent in existing approaches. Our method leverages an external timbre shifter during training to perturb the source speech's timbre, effectively mitigating timbre leakage and aligning the training process with the inference scenario. Additionally, we employed a diffusion transformer architecture that utilizes both a global timbre vector and prompt acoustic features, enhancing the model's ability to capture fine-grained timbre characteristics through in-context learning.

\textbf{Key Contributions}:
\begin{itemize}
    \item \textbf{Mitigating Timbre Leakage and Training-Inference Gap:} By introducing an external timbre shifter (OpenVoice) during training, we ensured that the semantic features extracted from the shifted utterance contain minimal residual timbre information from the source speaker. This approach effectively reduces timbre leakage and aligns the training objectives with the inference tasks, where content and timbre originate from different speakers.
    \item \textbf{Enhanced Timbre Representation:} Our dual representation strategy combines a timbre vector extracted from a speaker verification model (CAM++) with prompt acoustic features. This allows the model to access both global and fine-grained timbre information, improving its ability to generalize to unseen speakers in zero-shot scenarios.
    \item \textbf{Diffusion Transformer Architecture:} We adopted a diffusion transformer model with a flow matching diffusion scheme, enabling efficient denoising of acoustic features conditioned on semantic and timbre inputs. The transformer's capacity for in-context learning facilitates the utilization of the entire reference speech context, capturing nuanced timbre attributes essential for high-quality voice conversion.
    \item \textbf{Extension to Zero-Shot Singing Voice Conversion:} We extended Seed-VC to handle zero-shot singing voice conversion by incorporating fundamental frequency (F0) conditioning and gender-based pitch shift adjustments. This adaptation allows the model to preserve the expressive and melodic qualities of singing voices while adapting the timbre to unseen target singers.
\end{itemize}
\textbf{Experimental Validation}:
Our experiments demonstrated that Seed-VC outperforms strong baseline models (OpenVoice and CosyVoice) in zero-shot voice conversion tasks, achieving higher speaker similarity scores and lower word and character error rates.

In the context of zero-shot singing voice conversion, Seed-VC showed promising results, achieving higher speaker similarity and lower character error rates than the RVCv2 baseline while maintaining competitive F0 correlation and speech quality metrics. These findings highlight the versatility and effectiveness of Seed-VC in handling both spoken and singing voice conversion tasks without prior exposure to the target speakers.

\textbf{Future work}:
While the current study provides compelling evidence of Seed-VC's capabilities, there are areas for further exploration:
\begin{itemize}
    \item \textbf{Ablation Studies:} We plan to conduct comprehensive ablation studies to quantify the impact of different timbre shifter methods.
    \item \textbf{Dataset Expansion and Fair Comparison:} Future work will involve training Seed-VC on datasets identical to those used in previous zero-shot voice conversion studies to ensure fair comparisons. This will help validate the generalizability of our approach across different data distributions and speaker populations.
    \item \textbf{Real-Time Conversion and Efficiency Improvements:} Investigating techniques to reduce computational complexity and latency will be essential for deploying Seed-VC in real-time applications, such as live voice conversion and assistive technologies.
\end{itemize}

\bibliographystyle{unsrt}  


\begin{thebibliography}{10}
\providecommand{\url}[1]{#1}
\csname url@samestyle\endcsname
\providecommand{\newblock}{\relax}
\providecommand{\bibinfo}[2]{#2}
\providecommand{\BIBentrySTDinterwordspacing}{\spaceskip=0pt\relax}
\providecommand{\BIBentryALTinterwordstretchfactor}{4}
\providecommand{\BIBentryALTinterwordspacing}{\spaceskip=\fontdimen2\font plus
\BIBentryALTinterwordstretchfactor\fontdimen3\font minus \fontdimen4\font\relax}
\providecommand{\BIBforeignlanguage}[2]{{%
\expandafter\ifx\csname l@#1\endcsname\relax
\typeout{** WARNING: IEEEtran.bst: No hyphenation pattern has been}%
\typeout{** loaded for the language `#1'. Using the pattern for}%
\typeout{** the default language instead.}%
\else
\language=\csname l@#1\endcsname
\fi
#2}}
\providecommand{\BIBdecl}{\relax}
\BIBdecl

\bibitem{hsu2021hubert}
W.-N. Hsu, B.~Bolte, Y.-H.~H. Tsai, K.~Lakhotia, R.~Salakhutdinov, and A.~Mohamed, ``Hubert: Self-supervised speech representation learning by masked prediction of hidden units,'' \emph{IEEE/ACM transactions on audio, speech, and language processing}, vol.~29, pp. 3451--3460, 2021.

\bibitem{baevski2020wav2vec}
A.~Baevski, Y.~Zhou, A.~Mohamed, and M.~Auli, ``wav2vec 2.0: A framework for self-supervised learning of speech representations,'' \emph{Advances in neural information processing systems}, vol.~33, pp. 12\,449--12\,460, 2020.

\bibitem{sun2016phonetic}
L.~Sun, K.~Li, H.~Wang, S.~Kang, and H.~Meng, ``Phonetic posteriorgrams for many-to-one voice conversion without parallel data training,'' in \emph{2016 IEEE International Conference on Multimedia and Expo (ICME)}.\hskip 1em plus 0.5em minus 0.4em\relax IEEE, 2016, pp. 1--6.

\bibitem{qian2019autovc}
K.~Qian, Y.~Zhang, S.~Chang, X.~Yang, and M.~Hasegawa-Johnson, ``Autovc: Zero-shot voice style transfer with only autoencoder loss,'' in \emph{International Conference on Machine Learning}.\hskip 1em plus 0.5em minus 0.4em\relax PMLR, 2019, pp. 5210--5219.

\bibitem{van2017neural}
A.~Van Den~Oord, O.~Vinyals \emph{et~al.}, ``Neural discrete representation learning,'' \emph{Advances in neural information processing systems}, vol.~30, 2017.

\bibitem{baas2023voice}
M.~Baas, B.~van Niekerk, and H.~Kamper, ``Voice conversion with just nearest neighbors,'' \emph{arXiv preprint arXiv:2305.18975}, 2023.

\bibitem{ho2020denoising}
J.~Ho, A.~Jain, and P.~Abbeel, ``Denoising diffusion probabilistic models,'' \emph{Advances in neural information processing systems}, vol.~33, pp. 6840--6851, 2020.

\bibitem{kong2020diffwave}
Z.~Kong, W.~Ping, J.~Huang, K.~Zhao, and B.~Catanzaro, ``Diffwave: A versatile diffusion model for audio synthesis,'' \emph{arXiv preprint arXiv:2009.09761}, 2020.

\bibitem{chen2020wavegrad}
N.~Chen, Y.~Zhang, H.~Zen, R.~J. Weiss, M.~Norouzi, and W.~Chan, ``Wavegrad: Estimating gradients for waveform generation,'' \emph{arXiv preprint arXiv:2009.00713}, 2020.

\bibitem{popov2021diffusion}
V.~Popov, I.~Vovk, V.~Gogoryan, T.~Sadekova, M.~Kudinov, and J.~Wei, ``Diffusion-based voice conversion with fast maximum likelihood sampling scheme,'' \emph{arXiv preprint arXiv:2109.13821}, 2021.

\bibitem{vaswani2017attention}
A.~Vaswani, ``Attention is all you need,'' \emph{Advances in Neural Information Processing Systems}, 2017.

\bibitem{liu2022diffsinger}
J.~Liu, C.~Li, Y.~Ren, F.~Chen, and Z.~Zhao, ``Diffsinger: Singing voice synthesis via shallow diffusion mechanism,'' in \emph{Proceedings of the AAAI conference on artificial intelligence}, vol.~36, no.~10, 2022, pp. 11\,020--11\,028.

\bibitem{yang2024streamvc}
Y.~Yang, Y.~Kartynnik, Y.~Li, J.~Tang, X.~Li, G.~Sung, and M.~Grundmann, ``Streamvc: Real-time low-latency voice conversion,'' in \emph{ICASSP 2024-2024 IEEE International Conference on Acoustics, Speech and Signal Processing (ICASSP)}.\hskip 1em plus 0.5em minus 0.4em\relax IEEE, 2024, pp. 11\,016--11\,020.

\bibitem{choi2023diff}
H.-Y. Choi, S.-H. Lee, and S.-W. Lee, ``Diff-hiervc: Diffusion-based hierarchical voice conversion with robust pitch generation and masked prior for zero-shot speaker adaptation,'' \emph{International Speech Communication Association}, pp. 2283--2287, 2023.

\bibitem{huang2024mullivc}
J.~Huang, C.~Zhang, Y.~Ren, Z.~Jiang, Z.~Ye, J.~Liu, J.~He, X.~Yin, and Z.~Zhao, ``Mullivc: Multi-lingual voice conversion with cycle consistency,'' \emph{arXiv preprint arXiv:2408.04708}, 2024.

\bibitem{choi2024dddm}
H.-Y. Choi, S.-H. Lee, and S.-W. Lee, ``Dddm-vc: Decoupled denoising diffusion models with disentangled representation and prior mixup for verified robust voice conversion,'' in \emph{Proceedings of the AAAI Conference on Artificial Intelligence}, vol.~38, no.~16, 2024, pp. 17\,862--17\,870.

\bibitem{li2021ppg}
Z.~Li, B.~Tang, X.~Yin, Y.~Wan, L.~Xu, C.~Shen, and Z.~Ma, ``Ppg-based singing voice conversion with adversarial representation learning,'' in \emph{ICASSP 2021-2021 IEEE International Conference on Acoustics, Speech and Signal Processing (ICASSP)}.\hskip 1em plus 0.5em minus 0.4em\relax IEEE, 2021, pp. 7073--7077.

\bibitem{zhang2023leveraging}
X.~Zhang, Y.~Gu, H.~Chen, Z.~Fang, L.~Zou, L.~Xue, and Z.~Wu, ``Leveraging content-based features from multiple acoustic models for singing voice conversion,'' \emph{arXiv preprint arXiv:2310.11160}, 2023.

\bibitem{polyak2021speech}
A.~Polyak, Y.~Adi, J.~Copet, E.~Kharitonov, K.~Lakhotia, W.-N. Hsu, A.~Mohamed, and E.~Dupoux, ``Speech resynthesis from discrete disentangled self-supervised representations,'' \emph{arXiv preprint arXiv:2104.00355}, 2021.

\bibitem{li2023freevc}
J.~Li, W.~Tu, and L.~Xiao, ``Freevc: Towards high-quality text-free one-shot voice conversion,'' in \emph{ICASSP 2023-2023 IEEE International Conference on Acoustics, Speech and Signal Processing (ICASSP)}.\hskip 1em plus 0.5em minus 0.4em\relax IEEE, 2023, pp. 1--5.

\bibitem{peebles2023scalable}
W.~Peebles and S.~Xie, ``Scalable diffusion models with transformers,'' in \emph{Proceedings of the IEEE/CVF International Conference on Computer Vision}, 2023, pp. 4195--4205.

\bibitem{wang2023neural}
C.~Wang, S.~Chen, Y.~Wu, Z.~Zhang, L.~Zhou, S.~Liu, Z.~Chen, Y.~Liu, H.~Wang, J.~Li \emph{et~al.}, ``Neural codec language models are zero-shot text to speech synthesizers,'' \emph{arXiv preprint arXiv:2301.02111}, 2023.

\bibitem{lee2024ditto}
K.~Lee, D.~W. Kim, J.~Kim, and J.~Cho, ``Ditto-tts: Efficient and scalable zero-shot text-to-speech with diffusion transformer,'' \emph{arXiv preprint arXiv:2406.11427}, 2024.

\bibitem{eskimez2024e2}
S.~E. Eskimez, X.~Wang, M.~Thakker, C.~Li, C.-H. Tsai, Z.~Xiao, H.~Yang, Z.~Zhu, M.~Tang, X.~Tan \emph{et~al.}, ``E2 tts: Embarrassingly easy fully non-autoregressive zero-shot tts,'' \emph{arXiv preprint arXiv:2406.18009}, 2024.

\bibitem{chen2024f5}
Y.~Chen, Z.~Niu, Z.~Ma, K.~Deng, C.~Wang, J.~Zhao, K.~Yu, and X.~Chen, ``F5-tts: A fairytaler that fakes fluent and faithful speech with flow matching,'' \emph{arXiv preprint arXiv:2410.06885}, 2024.

\bibitem{qin2023openvoice}
Z.~Qin, W.~Zhao, X.~Yu, and X.~Sun, ``Openvoice: Versatile instant voice cloning,'' \emph{arXiv preprint arXiv:2312.01479}, 2023.

\bibitem{du2024cosyvoice}
Z.~Du, Q.~Chen, S.~Zhang, K.~Hu, H.~Lu, Y.~Yang, H.~Hu, S.~Zheng, Y.~Gu, Z.~Ma \emph{et~al.}, ``Cosyvoice: A scalable multilingual zero-shot text-to-speech synthesizer based on supervised semantic tokens,'' \emph{arXiv preprint arXiv:2407.05407}, 2024.

\bibitem{shen2023naturalspeech}
K.~Shen, Z.~Ju, X.~Tan, Y.~Liu, Y.~Leng, L.~He, T.~Qin, S.~Zhao, and J.~Bian, ``Naturalspeech 2: Latent diffusion models are natural and zero-shot speech and singing synthesizers,'' \emph{arXiv preprint arXiv:2304.09116}, 2023.

\bibitem{peng2024voicecraft}
P.~Peng, P.-Y. Huang, S.-W. Li, A.~Mohamed, and D.~Harwath, ``Voicecraft: Zero-shot speech editing and text-to-speech in the wild,'' \emph{arXiv preprint arXiv:2403.16973}, 2024.

\bibitem{bao2023all}
F.~Bao, S.~Nie, K.~Xue, Y.~Cao, C.~Li, H.~Su, and J.~Zhu, ``All are worth words: A vit backbone for diffusion models,'' in \emph{Proceedings of the IEEE/CVF conference on computer vision and pattern recognition}, 2023, pp. 22\,669--22\,679.

\bibitem{ronneberger2015u}
O.~Ronneberger, P.~Fischer, and T.~Brox, ``U-net: Convolutional networks for biomedical image segmentation,'' in \emph{Medical image computing and computer-assisted intervention--MICCAI 2015: 18th international conference, Munich, Germany, October 5-9, 2015, proceedings, part III 18}.\hskip 1em plus 0.5em minus 0.4em\relax Springer, 2015, pp. 234--241.

\bibitem{su2023enhanced}
J.~Su, Y.~Lu, S.~Pan, A.~Murtadha, B.~Wen, and Y.~L. Roformer, ``Enhanced transformer with rotary position embedding., 2021,'' \emph{DOI: https://doi. org/10.1016/j. neucom}, 2023.

\bibitem{wang2023cam++}
H.~Wang, S.~Zheng, Y.~Chen, L.~Cheng, and Q.~Chen, ``Cam++: A fast and efficient network for speaker verification using context-aware masking,'' \emph{arXiv preprint arXiv:2303.00332}, 2023.

\bibitem{casanova2022yourtts}
E.~Casanova, J.~Weber, C.~D. Shulby, A.~C. Junior, E.~G{\"o}lge, and M.~A. Ponti, ``Yourtts: Towards zero-shot multi-speaker tts and zero-shot voice conversion for everyone,'' in \emph{International Conference on Machine Learning}.\hskip 1em plus 0.5em minus 0.4em\relax PMLR, 2022, pp. 2709--2720.

\bibitem{he2024emilia}
H.~He, Z.~Shang, C.~Wang, X.~Li, Y.~Gu, H.~Hua, L.~Liu, C.~Yang, J.~Li, P.~Shi \emph{et~al.}, ``Emilia: An extensive, multilingual, and diverse speech dataset for large-scale speech generation,'' \emph{arXiv preprint arXiv:2407.05361}, 2024.

\bibitem{reddy2021dnsmos}
C.~K. Reddy, V.~Gopal, and R.~Cutler, ``Dnsmos: A non-intrusive perceptual objective speech quality metric to evaluate noise suppressors,'' in \emph{ICASSP 2021-2021 IEEE International Conference on Acoustics, Speech and Signal Processing (ICASSP)}.\hskip 1em plus 0.5em minus 0.4em\relax IEEE, 2021, pp. 6493--6497.

\bibitem{radford2023robust}
A.~Radford, J.~W. Kim, T.~Xu, G.~Brockman, C.~McLeavey, and I.~Sutskever, ``Robust speech recognition via large-scale weak supervision,'' in \emph{International conference on machine learning}.\hskip 1em plus 0.5em minus 0.4em\relax PMLR, 2023, pp. 28\,492--28\,518.

\bibitem{lee2022bigvgan}
S.-g. Lee, W.~Ping, B.~Ginsburg, B.~Catanzaro, and S.~Yoon, ``Bigvgan: A universal neural vocoder with large-scale training,'' \emph{arXiv preprint arXiv:2206.04658}, 2022.

\bibitem{zhang2022m4singer}
L.~Zhang, R.~Li, S.~Wang, L.~Deng, J.~Liu, Y.~Ren, J.~He, R.~Huang, J.~Zhu, X.~Chen \emph{et~al.}, ``M4singer: A multi-style, multi-singer and musical score provided mandarin singing corpus,'' \emph{Advances in Neural Information Processing Systems}, vol.~35, pp. 6914--6926, 2022.

\bibitem{wei2023rmvpe}
H.~Wei, X.~Cao, T.~Dan, and Y.~Chen, ``Rmvpe: A robust model for vocal pitch estimation in polyphonic music,'' \emph{arXiv preprint arXiv:2306.15412}, 2023.

\bibitem{kim2021conditional}
J.~Kim, J.~Kong, and J.~Son, ``Conditional variational autoencoder with adversarial learning for end-to-end text-to-speech,'' in \emph{International Conference on Machine Learning}.\hskip 1em plus 0.5em minus 0.4em\relax PMLR, 2021, pp. 5530--5540.

\bibitem{speechteam2024funaudiollm}
T.~SpeechTeam, ``Funaudiollm: Voice understanding and generation foundation models for natural interaction between humans and llms,'' \emph{arXiv preprint arXiv:2407.04051}, 2024.

\end{thebibliography}

\end{document}